\documentclass[aps,prb,reprint,superscriptaddress]{revtex4-2}
\usepackage[dvipdfmx]{graphicx}
\usepackage{color}
\usepackage{ulem}
\begin{document}
\title{Photoemission angular distribution beyond the single wavevector description of photoelectron final states}

\author{Hiroaki~Tanaka}
\email{hhiiro2022@gmail.com}
\affiliation{Institute for Solid State Physics, The University of Tokyo, Kashiwa, Chiba 277-8581, Japan}

\author{Shota~Okazaki}
\affiliation{Laboratory for Materials and Structures, Tokyo Institute of Technology, Yokohama, Kanagawa 226-8503, Japan}

\author{Yuto~Fukushima}
\affiliation{Institute for Solid State Physics, The University of Tokyo, Kashiwa, Chiba 277-8581, Japan}

\author{Kaishu~Kawaguchi}
\affiliation{Institute for Solid State Physics, The University of Tokyo, Kashiwa, Chiba 277-8581, Japan}

\author{Ayumi~Harasawa}
\affiliation{Institute for Solid State Physics, The University of Tokyo, Kashiwa, Chiba 277-8581, Japan}

\author{Takushi~Iimori}
\affiliation{Institute for Solid State Physics, The University of Tokyo, Kashiwa, Chiba 277-8581, Japan}

\author{Fumio~Komori}
\affiliation{Institute of Industrial Science, The University of Tokyo, Meguro-ku, Tokyo 153-8505, Japan}

\author{Masashi~Arita}
\affiliation{Hiroshima Synchrotron Radiation Center, Hiroshima University, Higashi-hiroshima, Hiroshima 739-0046, Japan}

\author{Ryo~Mori}
\affiliation{Institute for Solid State Physics, The University of Tokyo, Kashiwa, Chiba 277-8581, Japan}

\author{Kenta~Kuroda}
\email{kuroken224@hiroshima-u.ac.jp}
\affiliation{Graduate School of Advanced Science and Engineering, Hiroshima University, Higashi-hiroshima, Hiroshima 739-8526, Japan}
\affiliation{International Institute for Sustainability with Knotted Chiral Meta Matter (WPI-SKCM${}^{2}$), Hiroshima University, Higashi-hiroshima, Hiroshima 739-8526, Japan}
\affiliation{Research Institute for Semiconductor Engineering, Hiroshima University, Higashi-hiroshima, Hiroshima 739-8527, Japan}

\author{Takao~Sasagawa}
\email{sasagawa@msl.titech.ac.jp}
\affiliation{Laboratory for Materials and Structures, Tokyo Institute of Technology, Yokohama, Kanagawa 226-8503, Japan}
\affiliation{Research Center for Autonomous Systems Materialogy, Tokyo Institute of Technology, Yokohama, Kanagawa 226-8503, Japan}

\author{Takeshi~Kondo}
\email{kondo1215@issp.u-tokyo.ac.jp}
\affiliation{Institute for Solid State Physics, The University of Tokyo, Kashiwa, Chiba 277-8581, Japan}
\affiliation{Trans-scale Quantum Science Institute, The University of Tokyo, Bunkyo-ku, Tokyo 113-0033, Japan}

\date{\today}

\begin{abstract}
We develop a simulation procedure for angle-resolved photoemission spectroscopy (ARPES), where a photoelectron wave function is set to be an outgoing plane wave in a vacuum associated with the emitted photoelectron wave packet.
ARPES measurements on the transition metal dichalcogenide $1T$-$\mathrm{Ti}\mathrm{S}_2$ are performed, and our simulations exhibit good agreement with experiments.
Analysis of our calculated final state wave functions quantitatively visualizes that they include various waves due to the boundary condition and the uneven crystal potential.
These results show that a more detailed investigation of the photoelectron final states is necessary to fully explain the photon-energy- and light-polarization-dependent ARPES spectra.
\end{abstract}

\maketitle


\newcommand{\para}{{/\mkern-5mu/}}

Angle-resolved photoemission spectroscopy (ARPES) has been a powerful method for investigating the electronic structure of solid crystals \cite{RevModPhys.93.025006}.
It is based on the photoelectric effect \cite{https://doi.org/10.1002/andp.18872670827}; the energies and momenta of photoelectrons derive the band dispersion of a solid.
As well as the band dispersion, the intensity distribution of ARPES spectra gives information on the wave function of each band because the matrix element between the ground state and the photoexcited state determines it.
Particularly, the light polarization dependence of ARPES spectra has extracted the electronic structure information of various kinds of materials \cite{Day2019, PhysRevX.9.041049, PhysRevLett.107.207602, PhysRevB.84.125422, PhysRevLett.107.166803, doi:10.1126/sciadv.aay2730}.

\begin{figure}
\includegraphics{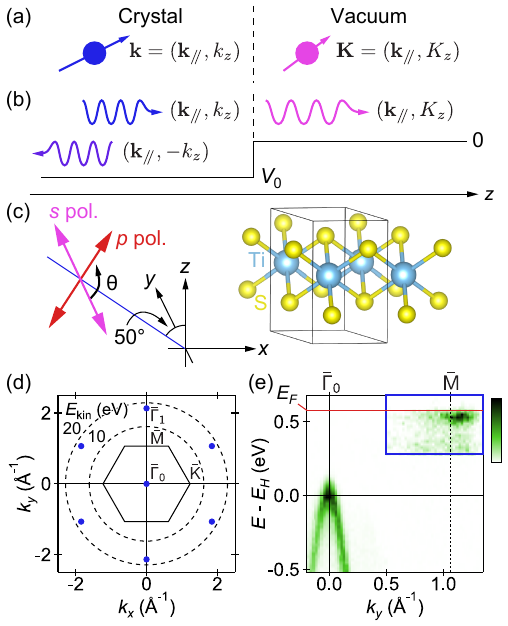}
\caption{\label{Fig: Fundamental} Description of the photoemission process and fundamental properties of $1T$-$\mathrm{Ti}\mathrm{S}_2$. (a) Three-step model description of the photoemission. (b) Schematic of a photoelectron wave function discussed in the present Letter. (c) Crystal structure and incident light direction in ARPES measurements. (d) Brillouin zone (black hexagon) and reciprocal lattice points (blue dots). Black dashed circles represent the wavevector length satisfying $E_\mathrm{kin}=\hbar^2k^2/2m$. (e) Experimental band dispersion along the $k_y$ direction taken by 35 eV synchrotron light.}
\end{figure}

In response to growing interest in the photoemission angular distribution, numerical studies have reported its simulation methodologies \cite{Day2019, PhysRevB.51.13614, DAIMON1995487, Nishimoto_1996, NISHIMOTO1996671, PhysRevB.56.7687, MOSER201729, TANAKA2023147297}.
We focus on calculating a photoemission matrix element derived from Fermi's golden rule; on the other hand, there has been another approach to ARPES spectra simulation by time-dependent density functional theory \cite{doi:10.1021/acs.jctc.6b00897, PhysRevResearch.5.033075}.
A plane wave is the simplest form for describing final states, and some studies have added the scattering effects for better reproducibility \cite{PhysRevB.83.115437, KRUGER2022147219, PhysRevResearch.5.033075}.
These studies have assumed that the photoelectron wave function in a solid is associated with a single wavevector $\mathbf{k}=(\mathbf{k}_\para, k_z)$ [Fig.\ \ref{Fig: Fundamental}(a)]; $\mathbf{k}_\para$ represents the in-plane ($xy$) components.
This assumption is consistent with the three-step model of photoemission \cite{Hufner2003}, where a photoelectron is described as a classical particle with a certain momentum.
On the other hand, the time evolution analysis of photoemission has revealed that a photoelectron propagates as a wave packet, a cropped plane wave, in a vacuum \cite{HiroakiTanaka20232023-013}.
The wave-like behavior of photoelectrons tells us that time-reversed low-energy electron diffraction (TR-LEED) states, which have been employed for decades \cite{PhysRev.134.A788, PhysRevB.2.4334}, are appropriate for analyzing the whole amplitude of photoelectron wave packet [see Note 1 in Supplemental Material (SM) \cite{Supplemental}]; The LEED states are determined by the boundary condition that there is only the transmitted wave in the $z\rightarrow-\infty$ limit and the TR-LEED states are obtained by taking the time reversal of them.
Since TR-LEED states are determined by the boundary condition at the vacuum layer below the slab, one needs to calculate the TR-LEED wave function for the entire range of the thick slab, leading to a large numerical error in the near-surface region, which is crucial for analyzing surface-sensitive photoemission.
One method to avoid this difficulty is to add an imaginary constant optical potential, making the photoelectron wave function rapidly decay into bulk \cite{PhysRevLett.93.027601, PhysRevB.74.195125, PhysRevB.75.045432, PhysRevLett.129.246404, Strocov2023}.
However, a recent ARPES study on layered materials has reported that the exponential decay associated with the constant optical potential cannot reproduce the ARPES spectra modulation by the surface sensitivity \cite{PhysRevLett.132.136402}.
\nocite{PhysRevB.5.4709} 
\nocite{doi:10.1021/acs.jpcc.2c08140}
\nocite{ITC}
\nocite{PhysRevB.67.155108}
\nocite{Giannozzi2009}
\nocite{Giannozzi2017}
\nocite{Pseudopotentials}
\nocite{PhysRevLett.77.3865, PhysRevLett.78.1396}
\nocite{doi:10.1063/1.1564060}
\nocite{doi:10.1063/1.2204597}
\nocite{Pizzi_2020}
\nocite{doi:10.1021/acs.inorgchem.8b02883}
\nocite{Marsman_2008}

In this Letter, we propose another approach to the photoemission simulation within the Kohn-Sham system of the density functional theory \cite{PhysRev.136.B864, PhysRev.140.A1133} by approximating the photoelectron wave function to include only an outgoing plane wave in the vacuum above the slab [pink curve in Fig.\ \ref{Fig: Fundamental}(b)].
This approximation removing the ingoing plane waves is supported by our one-dimensional photoemission simulations (Fig.\ S3 in SM \cite{Supplemental}).
In this situation, the photoelectron wave function inside a solid can contain various waves other than the traveling wave.
Even in the stepped potential model, the connection condition at the boundary requires the wave function to include the reflected wave $(\mathbf{k}_\para, -k_z)$  [purple curve in Fig.\ \ref{Fig: Fundamental}(b)].
Consequently, the photoemission angular distribution using such final-state wave functions can differ from that using the three-step model wave function, including only the traveling wave; we refer to them as first-principles (FP) and plane-wave (PW) final states, respectively.
Our photoemission intensity calculations were validated by comparing them with ARPES spectra of the transition-metal dichalcogenide (TMD) $1T$-$\mathrm{Ti}\mathrm{S}_2$ [Fig.\ \ref{Fig: Fundamental}(c)].
While the electronic structure of TMDs has been investigated by ARPES \cite{PhysRevLett.111.106801, Chen2015, PhysRevB.105.L121102, PhysRevB.105.245145}, $1T$-$\mathrm{Ti}\mathrm{S}_2$ is ideal for our study because of the easily cleavable quasi-two-dimensional structure, no need for considering the termination surface dependence \cite{PhysRevB.105.235126}, and the simple electronic structure without the charge density wave phase in contrast to other $1T$-type TMDs \cite{PhysRevLett.32.882}.
We found that the intensity distributions of FP final states were closer to experimental ones than those of PW final states.
Moreover, our wave component analysis of FP wave functions demonstrated that they include various plane waves beyond just the traveling wave; these components, absent in the PW final states, contributed to a better agreement between experiments and simulations.
Our study is a quantitative demonstration that the final state wave functions include reflected waves in a solid.
Our study provides a precise description of the photoemission process, which can be particularly important when extracting physical quantities from photoemission angular distribution.

Single crystals of $1T$-$\mathrm{Ti}\mathrm{S}_2$ were grown by the chemical vapor transport method with iodine as the transport agent.
Magnetotransport and x-ray photoemission measurements exhibited only trivial behavior, indicating the simple electronic structure of this material (Notes 2 and 3 in SM \cite{Supplemental}).
Laue back-reflection measurements determined the crystal geometry as described in Fig.\ \ref{Fig: Fundamental}(c) (Note 4 in SM \cite{Supplemental}).
ARPES measurements using a 7-eV laser and synchrotron radiation ($h\nu=9\text{--}39\ \mathrm{eV}$) were performed at ISSP, the University of Tokyo \cite{10.1063/1.4948738, 10.1063/5.0151859} and HiSOR BL-9A, respectively (Note 5 in SM \cite{Supplemental}).
$1T$-$\mathrm{Ti}\mathrm{S}_2$ possesses a hole band around the $\bar{\Gamma}$ point and an electron band around the $\bar{M}$ point, forming an indirect band gap [Fig.\ \ref{Fig: Fundamental}(e)].
We focus on the hole band in discussing the photoemission intensity; we represent its top position by $E_H$ and use it as the base point of the energy axis.

\begin{figure}
\includegraphics{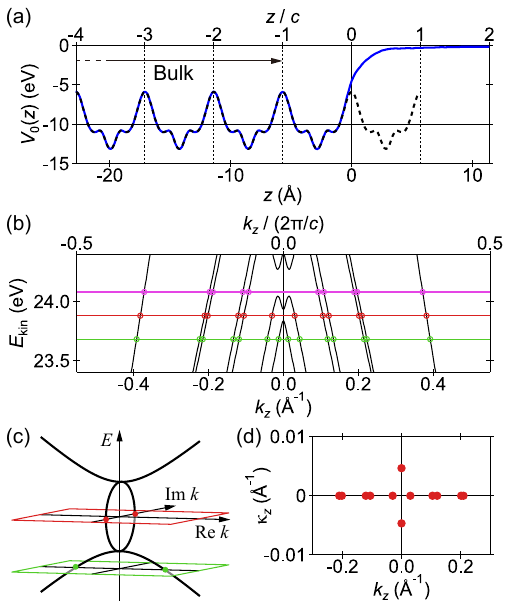}
\caption{\label{Fig: Methods} FP wave function calculations. (a) $\mathbf{g}_\para=0$ component of the Fourier-expanded potential. The $1T$-$\mathrm{Ti}\mathrm{S}_2$ slab is placed in the $z<0$ region. Black dashed curves and solid horizontal lines represent the periodic bulk potential and its $\mathbf{g}=0$ component. (b) Bulk band dispersion of unoccupied states along the real $k_z$ axis. The in-plane momentum is set to $(0.1, 0.0)\ (\text{\AA}^{-1})$. (c) Band dispersion of the effective model around the hybridization gap. Only the real eigenvalues are plotted. The constant-energy planes colored red and green correspond to the horizontal lines in (b). (d) Magnified view of the $z$ component distribution with the kinetic energy corresponding to the red line in (b).}
\end{figure}

Before presenting the experimental results, we explain the calculation procedure for the FP final states.
We implemented the FP wave function calculations in SPADExp \cite{TANAKA2023147297}, which uses OpenMX \cite{PhysRevB.67.155108} to calculate the ground state wave functions and the Kohn-Sham potential.
The ground state electronic structure for a 40-layer slab without surface relaxation (Note 7 in SM \cite{Supplemental}) was obtained using the Perdew-Burke-Ernzerhof (PBE) functional \cite{PhysRevLett.77.3865, PhysRevLett.78.1396}.
Although the system becomes semimetallic due to the band gap underestimation of the PBE functional (Note 6 in SM \cite{Supplemental}), the overall shape of the quasi-two-dimensional hole band around the $\bar{\Gamma}$ point agrees well with experimental results.
See Note 8 in SM \cite{Supplemental} for other calculation properties.

\begin{figure*}
\includegraphics{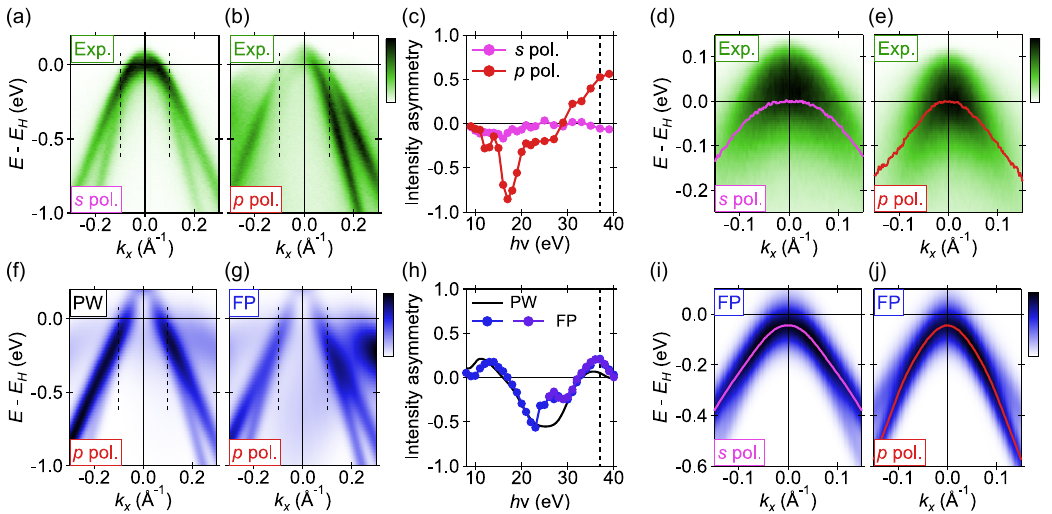}
\caption{\label{Fig: ARPES} Experimental ARPES spectra and simulation results. (a), (b) Experimental spectra along the $k_x$ direction taken by $s$- and $p$-polarized 37 eV synchrotron lights, respectively. (c) Photon energy dependence of the integrated intensity asymmetry along the dashed lines in (a) and (b). The vertical dashed line is located at $h\nu=37\ \mathrm{eV}$. (d), (e) ARPES spectra taken by the 7 eV laser. The overlaid curves represent peak positions extracted from energy distribution curves. (f), (g) Simulated photoemission angular distribution using the PW and FP final states. The incident light was 37 eV, $p$ polarized. (h) Integrated intensity asymmetry calculated from simulated spectra. The blue and purple curves represent the FP final-state calculations with $\mathbf{g}_\para$-limited and $\mathbf{g}_\para$-unlimited conditions, respectively. (i), (j) Simulated spectra corresponding to (d) and (e) using the FP final states, respectively.}
\end{figure*}

The FP wave functions are calculated using the local part of the Kohn-Sham potential and the kinetic energy value; the latter is determined by the binding energy of the ground state, the incident photon energy, and the work function.
Owing to the in-plane periodicity of the system, the wave function can be Fourier expanded,
\begin{equation}
\psi^{F}(\mathbf{r})=\sum_{\mathbf{g}_\para}  e^{i(\mathbf{k}_\para+\mathbf{g}_\para)\cdot\mathbf{r}_\para}\cdot \psi_{\mathbf{g}_\para}(z), \label{eq: slab}
\end{equation}
where $\mathbf{r}=(\mathbf{r}_\para, z)$ holds and $\mathbf{g}_\para$ represents an in-plane reciprocal lattice vector, corresponding to the blue dots in Fig.\ \ref{Fig: Fundamental}(d).
This representation is available both in a solid and a vacuum.
The Fourier-expanded potential is also periodic along the $z$ direction inside the crystal except for the top and bottom layers [Fig.\ \ref{Fig: Methods}(a)].
Therefore, the wave function in the region is the linear combination of the bulk wave function:
\begin{equation}
\psi^{F}(\mathbf{r})=\sum_n p_n \psi_n^\mathrm{bulk}(\mathbf{r}),\ 
\psi_n^\mathrm{bulk}(\mathbf{r})=\sum_\mathbf{g} e^{i(\mathbf{k}_n+\mathbf{g})\cdot\mathbf{r}}\cdot \psi_{n\mathbf{g}}, \label{eq: bulk}
\end{equation}
where $n$ is the band index, $\mathbf{g}=(\mathbf{g}_\para, g_z)$ is a bulk reciprocal vector, and $\psi_{n\mathbf{g}}$ is a number representing an eigenstate.
Equation (\ref{eq: bulk}) is available only in the bulk region.

The $z$ component of the wavevector $\mathbf{k}_n$ is determined so that the eigenenergy equals the specified kinetic energy and is not limited to a real value because of the incomplete translational symmetry of the slab system along the $z$ direction; $\mathbf{k}_n$ can be decomposed as $(\mathbf{k}_\para, k_{nz}-i\kappa_{nz})$.
The complex $z$ component is necessary when the specified kinetic energy is located within a hybridization gap of the unoccupied band dispersion [red and pink lines in Fig.\ \ref{Fig: Methods}(b)].
In these cases, the number of eigenstates on the real $k_z$ axis is fewer than otherwise [green line in Fig.\ \ref{Fig: Methods}(b)], making them insufficient as a basis set.
Considering the effective Hamiltonian around the hybridization gap $H(k)=\left( \begin{array}{cc} ak & t \\ t & -ak \end{array}\right)$ and extend $k$ into the complex plane, we can find two bands within the hybridization gap [Fig.\ \ref{Fig: Methods}(c)].
While two eigenstates are on the real $k_z$ axis when the specified kinetic energy is out of the gap [green dots in Fig.\ \ref{Fig: Methods}(c)], they are on the complex plane when the kinetic energy is within the gap [red dots in Fig.\ \ref{Fig: Methods}(c)].
Also in the case of the unoccupied states in Fig.\ \ref{Fig: Methods}(b), we find two eigenstates with complex $z$ components for each gapped dispersion pair [Figs.\ \ref{Fig: Methods}(d) and S10 in SM \cite{Supplemental}].
The other case happens when $\mathbf{g}_\para$ and $\kappa_z$ are relatively large, but it only weakly affects the FP wave function (Note 9 in SM \cite{Supplemental}).
The linear combination coefficient $p_n$ and the wave function of the top layer are determined to satisfy the Schr\"odinger equation and the boundary condition in a vacuum; $\psi_{\mathbf{g}_\para}(z)$ is equal to $e^{i K_z z}$ when $\mathbf{g}_\para=0$ and to zero otherwise.
After these calculations, the decay function is multiplied to effectively include the surface-sensitive property of photoemission spectroscopy (Ref.\ \cite{https://doi.org/10.1002/sia.740010103} and Note 8 in SM \cite{Supplemental}).
We note that the projector-type nonlocal component of pseudopotentials \cite{PhysRevB.47.6728}, which affects the wave function around the nuclei, is omitted in the FP wave function calculations for the stability and computational cost reduction (Note 10 in SM \cite{Supplemental}).

We applied the FP wave functions to simulate the photoemission angular distribution of $1T$-$\mathrm{Ti}\mathrm{S}_2$ and compared the results with experimental spectra.
Experimental ARPES spectra using synchrotron light strongly depended on the incident photon energy and polarization [Figs.\ \ref{Fig: ARPES}(a), \ref{Fig: ARPES}(b), S11, and S12 in SM \cite{Supplemental}], indicating that the photoemission matrix element strongly depends on those conditions.
The $s$-polarization spectra were symmetric, as both the crystal structure and the light electric field are symmetric with respect to the $yz$ plane, while $p$-polarized spectra were not.
We determined the degree of asymmetry $(I_+-I_-)/(I_++I_-)$, where $I_\pm$ represents the integrated photoemission intensity at $k_x=\pm0.1\ \text{\AA}^{-1}$ [dashed lines in Figs.\ \ref{Fig: ARPES}(a) and \ref{Fig: ARPES}(b)].
For the $s$-polarized light, the degree of asymmetry was nearly zero across the whole photon energy range.
On the other hand, the degree of asymmetry for the $p$-polarized light varied significantly with the photon energy [Fig.\ \ref{Fig: ARPES}(c)], while its momentum dependence was weak (Fig.\ S13 in SM \cite{Supplemental}).
This strongly varying behavior indicates that the degree of asymmetry is one proper quantity to discuss the photoemission matrix element.

Photoemission intensity simulations using the FP wave functions could reproduce similar trends as shown in Fig.\ \ref{Fig: ARPES}(h).
In the calculations, the light polarization affects the photoemission intensity as the vector potential direction in the matrix element.
We found that the aforementioned method for FP final states becomes unstable at the incident photon energies below 25 eV (Note 10 in SM \cite{Supplemental}).
In this case, we limited the $\mathbf{g}_\para$ vector within the circle determined by the kinetic energy [Fig.\ \ref{Fig: Fundamental}(d)] and solved the ordinal differential equation for the slab wave function [Eq.\ (\ref{eq: slab})].
The $\mathbf{g}_\para$-limited calculations [blue curve in Fig.\ \ref{Fig: ARPES}(h)] were also applicable for photon energies greater than 25 eV while the $\mathbf{g}_\para$-unlimited version [purple curve in Fig.\ \ref{Fig: ARPES}(h)] was suitable for the photon energy between 25 and 40 eV.
Both calculations exhibited very similar dispersions within this range, suggesting the large $\mathbf{g}_\para$ component is not so considerable.
Comparing the PW and FP final states, we claim that the FP final states exhibit better agreement with the sharp local minimum observed around $h\nu=20\ \mathrm{eV}$ and larger positive asymmetry in $h\nu=35\text{--}40\ \mathrm{eV}$.
The latter difference is clearly visualized in Figs.\ \ref{Fig: ARPES}(f) and \ref{Fig: ARPES}(g) for $h\nu=37\ \mathrm{eV}$, corresponding to Fig.\ \ref{Fig: ARPES}(b).

In laser ARPES measurements, we could investigate the polarization dependence in more detail by continuously changing the polarization angle $\theta$ in Fig.\ \ref{Fig: Fundamental}(c).
The $s$- and $p$-polarization spectra showed similar behavior to synchrotron ARPES measurements; the spectrum peaks for the $s$-polarization measurements appeared outside, while those for the $p$-polarization were inside [Figs.\ \ref{Fig: ARPES}(d) and \ref{Fig: ARPES}(e)].
Furthermore, we observed these peak positions oscillated continuously with the change in $\theta$ [Fig.\ S15(c) in SM \cite{Supplemental}].
These behaviors were also well reproduced in simulations using the FP final states [Figs.\ \ref{Fig: ARPES}(i), \ref{Fig: ARPES}(j), and S16(a)--S16(c) in SM \cite{Supplemental}].
In contrast, the PW final states exhibited worse agreement with the $p$-polarization spectra and the smoothness of the $\theta$ dependence curves [Figs.\ S16(d)--S16(f) in SM \cite{Supplemental}].
To summarize, the FP final states agreed better with both synchrotron and low-energy laser ARPES results than the PW final states.

\begin{figure}
\includegraphics{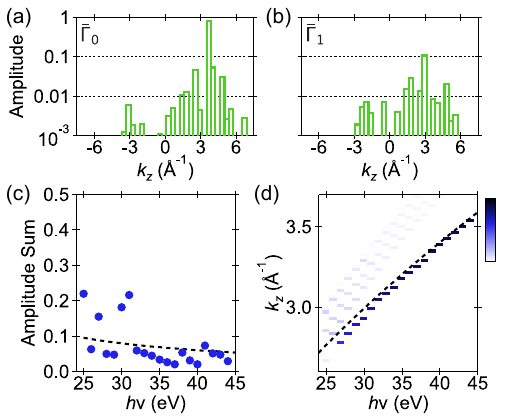}
\caption{\label{Fig: Components} Amplitude analysis of the FP wave function. (a), (b) Amplitude histogram for $\mathbf{g}_\para=\bar{\Gamma}_0, \bar{\Gamma}_1$. (c) Amplitude sum of the reflected ($k_z<0$) waves in (a) and its photon energy dependence. The dashed curve represents the theoretical value of the stepped potential model. (d) Photon energy dependence of the amplitude histogram around the prominent peak. The dashed curve represents the $k_z$ value of the three-step model.}
\end{figure}

Finally, we analyze the FP wave functions to quantitatively show that they include various plane waves.
For this analysis, we consider the case when the in-plane wavevector, the binding energy, and the excitation energy are $(0.1, 0.0)\ (\text{\AA}^{-1})$, 0 eV (Fermi level), and 37 eV respectively, corresponding to Fig.\ \ref{Fig: ARPES}(g).
From Eq.\ (\ref{eq: bulk}), the FP wave function includes waves with wavevector $\mathbf{k}_n+\mathbf{g}=(\mathbf{k}_\para+\mathbf{g}_\para)+(k_{nz}+g_z)-i\kappa_{nz}$ and amplitude $|p_n\psi_{n\mathbf{g}}|$.
Figures \ref{Fig: Components}(a) and \ref{Fig: Components}(b) show the histogram of these amplitudes for $\mathbf{g}_\para=\bar{\Gamma}_0, \bar{\Gamma}_1$ [defined in Fig.\ \ref{Fig: Fundamental}(d)] and $\kappa_{nz}=0$; $\kappa_{nz}\not=0$ waves have only negligible amplitudes (Fig.\ S17 in SM \cite{Supplemental}).
While a prominent peak may be associated with the photoelectron wavevector of the three-step model, other components with different $k_z$ values exist, each having amplitudes of at most 0.1.
Furthermore, the wave function includes components with $k_z<0$ (reflected waves) and $\mathbf{g}_\para=\bar{\Gamma}_1$ (different in-plane wavevector).
While the former is explained using the stepped potential model, as we discuss in the introduction, the latter is not and is due to the uneven crystal potential.
The amplitude for larger $\mathbf{g}_\para$ is at most 0.01 (Fig.\ S18 in SM \cite{Supplemental}), consistent with no large difference observed between the $\mathbf{g}_\para$-limited and $\mathbf{g}_\para$-unlimited calculations [Fig.\ \ref{Fig: ARPES}(h)].
For further analysis, Fig.\ \ref{Fig: Components}(c) compares the amplitude sum of the reflected waves and the theoretical value of the stepped potential model.
Our simulation and the stepped potential model show similar degrees of amplitudes and decreasing behavior with respect to the photon energy.
The prominent peak in the amplitude histogram can be compared with the $k_z$ value of the three-step model [Fig.\ \ref{Fig: Components}(d)].
A slight deviation is evident when the photon energy is around 30 eV, but it disappears around 40 eV.
Such behavior is attributed to the higher photoelectron kinetic energy, that renders the potential unevenness more negligible.
These results show that our FP wave functions share common characteristics with the stepped potential model and the three-step model, although they might seem too simple to reproduce the experimental ARPES spectra.
We note that this analysis does not consider the decay function related to the ARPES surface sensitivity.
While the disagreement between ARPES experiments and calculations has been reported \cite{PhysRevB.105.245145, PhysRevLett.132.136402}, they will be mainly affected by the decay function, particularly its length scale.

In conclusion, we develop a numerical method to calculate the FP final states from the local Kohn-Sham potential.
Leveraging the periodic property of the slab potential, we construct the wave function using the bulk eigenstates.
We performed ARPES measurements on the transition metal dichalcogenide $1T$-$\mathrm{Ti}\mathrm{S}_2$ and compared the photon-energy- and light-polarization-dependent results with simulations using our calculated FP final states and the PW final states.
We revealed that the FP final states include various plane waves such as reflected ($k_z<0$) waves and those with different in-plane wavevector ($\mathbf{g}_\para\not=0$), and they contributed to a better coincidence between experiments and simulations.
Our result can fill the gap between experiments and simplified simulations, which may connect the ARPES intensity distribution with the detail of the electronic structure, such as orbital components and the Berry phase.
While we demonstrate that our method is better than the plane-wave approximation, we hope that further research compares the availability of various photoemission simulations beyond the plane-wave approximation in more detail, including ours and one using TR-LEED states with the imaginary optical potential correction \cite{PhysRevLett.93.027601, PhysRevB.74.195125, PhysRevB.75.045432, PhysRevLett.129.246404, Strocov2023} or plane waves modified by the scattering \cite{PhysRevB.83.115437, KRUGER2022147219, PhysRevResearch.5.033075}.

\begin{acknowledgments}
This work was supported by Grant-in-Aid for JSPS Fellows (Grant No.\ JP21J20657), 
Grant-in-Aid for Early-Career Scientists (Grant No.\ JP23K13041),
Grant-in-Aid for Challenging Research (Pioneering) (Grants No.\ JP21K18181 and No.\ JP23K17351), 
Grant-in-Aid for Scientific Research (B) (Grant No.\ JP22H01943),
Grant-in-Aid for Transformative Research Areas (A) (Grant No.\ JP21H05236), 
Grant-in-Aid for Scientific Research (A) (Grants No.\ JP19H00651, No.\ JP21H04439, and No.\ JP21H04652), 
the Asahi Glass Foundation, 
the Murata Science Foundation, and 
Collaborative Research Projects of Laboratory for Materials and Structures, Tokyo Institute of Technology.
Laue back-reflection measurements were performed using the facilities of the Materials Design and Characterization Laboratory at the Institute for Solid State Physics, the University of Tokyo.
The synchrotron radiation experiments were performed with the approval of Hiroshima Synchrotron Radiation Center (HiSOR) (Proposal No.\ 22BG046).
\end{acknowledgments}

\bibliography{TiS2_ARPES_reference}
\end{document}


\title{Supplemental Material: Photoemission angular distribution beyond the single wavevector description of photoelectron final~states}

\author{Hiroaki~Tanaka}
\affiliation{Institute for Solid State Physics, The University of Tokyo, Kashiwa, Chiba 277-8581, Japan}

\author{Shota~Okazaki}
\affiliation{Laboratory for Materials and Structures, Tokyo Institute of Technology, Yokohama, Kanagawa 226-8503, Japan}

\author{Yuto~Fukushima}
\affiliation{Institute for Solid State Physics, The University of Tokyo, Kashiwa, Chiba 277-8581, Japan}

\author{Kaishu~Kawaguchi}
\affiliation{Institute for Solid State Physics, The University of Tokyo, Kashiwa, Chiba 277-8581, Japan}

\author{Ayumi~Harasawa}
\affiliation{Institute for Solid State Physics, The University of Tokyo, Kashiwa, Chiba 277-8581, Japan}

\author{Takushi~Iimori}
\affiliation{Institute for Solid State Physics, The University of Tokyo, Kashiwa, Chiba 277-8581, Japan}

\author{Fumio~Komori}
\affiliation{Institute of Industrial Science, The University of Tokyo, Meguro-ku, Tokyo 153-8505, Japan}

\author{Masashi~Arita}
\affiliation{Hiroshima Synchrotron Radiation Center, Hiroshima University, Higashi-hiroshima, Hiroshima 739-0046, Japan}

\author{Ryo~Mori}
\affiliation{Institute for Solid State Physics, The University of Tokyo, Kashiwa, Chiba 277-8581, Japan}

\author{Kenta~Kuroda}
\affiliation{Graduate School of Advanced Science and Engineering, Hiroshima University, Higashi-hiroshima, Hiroshima 739-8526, Japan}
\affiliation{International Institute for Sustainability with Knotted Chiral Meta Matter (WPI-SKCM${}^{2}$), Hiroshima University, Higashi-hiroshima, Hiroshima 739-8526, Japan}
\affiliation{Research Institute for Semiconductor Engineering, Hiroshima University, Higashi-hiroshima, Hiroshima 739-8527, Japan}

\author{Takao~Sasagawa}
\affiliation{Laboratory for Materials and Structures, Tokyo Institute of Technology, Yokohama, Kanagawa 226-8503, Japan}
\affiliation{Research Center for Autonomous Systems Materialogy, Tokyo Institute of Technology, Yokohama, Kanagawa 226-8503, Japan}

\author{Takeshi~Kondo}
\affiliation{Institute for Solid State Physics, The University of Tokyo, Kashiwa, Chiba 277-8581, Japan}
\affiliation{Trans-scale Quantum Science Institute, The University of Tokyo, Bunkyo-ku, Tokyo 113-0033, Japan}

\date{\today}

\maketitle

\clearpage

\renewcommand{\thefigure}{S\arabic{figure}}
\renewcommand{\thetable}{S\Roman{table}}
\titleformat*{\section}{\bfseries}
\newcommand{\para}{{/\mkern-5mu/}}

\section{Note 1: Description of photoelectron final states}
\begin{figure}[hb]
\includegraphics{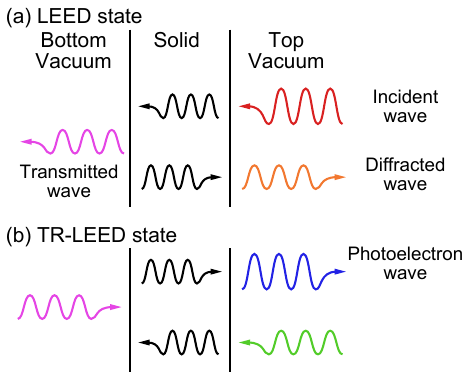}
\caption{\label{Fig: Ph_wfns} Schematics of LEED and TR-LEED states.}
\end{figure}

Here, we discuss the difference between a photoelectron wave function used in our present study and one in previous studies.
We consider a one-dimensional thick slab system with vacuum layers on top and bottom [Fig.\ \ref{Fig: Ph_wfns}] for simple correspondence with calculations.


First, we discuss the time-reversed low-energy electron diffraction (TR-LEED) state [Fig.\ \ref{Fig: Ph_wfns}(b)].
The TR-LEED state has been employed for decades, from photocurrent studies [22, 23] to recent ones simulating ARPES spectra [25--29].
A LEED state comprises incident and diffracted waves in the top vacuum, a Bloch wave in the slab, and a transmitted wave in the bottom vacuum [Fig.\ \ref{Fig: Ph_wfns}(a)].
The boundary condition is that there is only a transmitted wave in the bottom vacuum layer.
Therefore, the schematic in Fig.\ \ref{Fig: Ph_wfns}(b) represents the TR-LEED state.
The TR-LEED state is appropriate for analyzing the photoemission process because it includes the whole amplitude of the photoelectron wave packet.
When the photoelectron kinetic energy is specified, there are a pair of orthogonalized wave functions: in the case of a free electron, the pair is like $\exp(ikx)$ and $\exp(-ikx)$, or $\sin(kx)$ and $\cos(kx)$.
Therefore, the appropriate photoelectron wave function satisfies that the other wave function doesn't include the outgoing plane wave in the top vacuum layer related to the photoelectron wave packet, and the TR-LEED state firmly satisfies such a condition.

\begin{figure}
\includegraphics{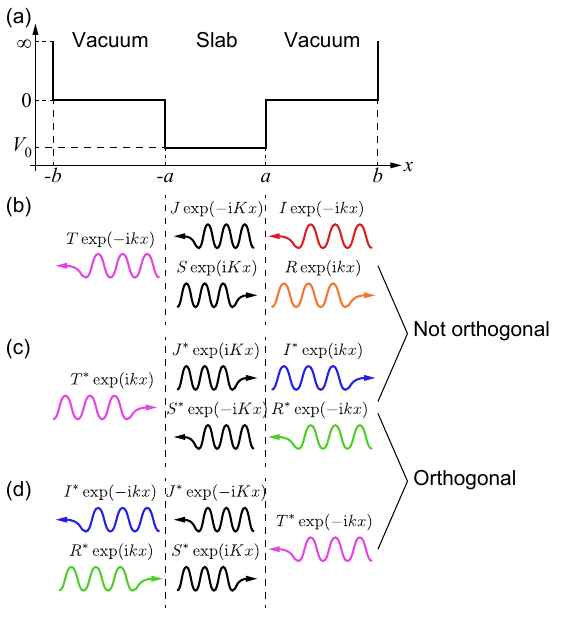}
\caption{\label{Fig: Ph_well} Schematics of unbound state wave functions. (a) Quantum well potential model. The fixed-end boundaries at $x=\pm b$ are necessary for numerical simulations. (b), (c) LEED and TR-LEED states. The wavevectors $k$ and $K$ satisfy $k=\sqrt{2E},\ K=\sqrt{2(E-V_0)}$, where $E$ is the kinetic energy of a photoelectron. These wave functions can be interpreted as single-wave states if we see the $x<0$ region. (d) Space-inverted TR-LEED state.}
\end{figure}

We explain how the TR-LEED state satisfies the requirement for photoelectron wave function using the quantum well potential model [Fig.\ \ref{Fig: Ph_well}(a)].
In this case, the TR-LEED state and space-inverted TR-LEED state [Figs.\ \ref{Fig: Ph_well}(c) and \ref{Fig: Ph_well}(d) respectively] form a pair, and the latter does not include the outgoing plane wave in the $x>a$ region; the orthogonality relation holds because $R/T=\mathrm{i}(K^2-k^2)\sin(2Ka)/2kK$ is a pure imaginary number and therefore the contributions to the inner product of them from top and bottom vacuum layers cancel out.
While the incoming plane wave in the TR-LEED state seems inconsistent with photoemission experiments, it is unimportant because it does not form any photoelectron wave packet.
If we remove it, the single-wave wave function becomes inappropriate for photoemission analysis.
When we see the $x<0$ side, the LEED state [Fig.\ \ref{Fig: Ph_well}(b)] can be interpreted as the single-wave state.
While the time-reversal of it [Fig.\ \ref{Fig: Ph_well}(c)] does not include the outgoing plane wave in the $x<-a$ region, they are not orthogonal.

\begin{figure}
\includegraphics[width=100mm]{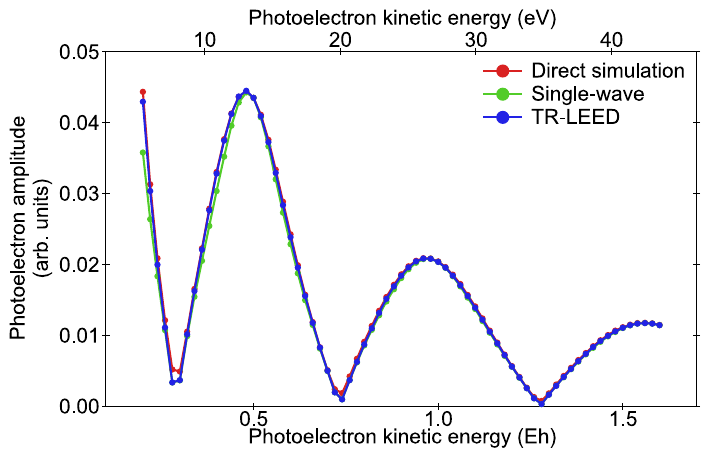}
\caption{\label{Fig: Photoemission_1d} Simulated photoelectron amplitude from the one-dimensional well potential system [Fig.\ \ref{Fig: Ph_well}(a)]. Parameters used for the simulations are $a=10,\ b=1000,\ V_0=-0.5,\ t=100/k$.}
\end{figure}

The above discussion on the photoelectron wave function is also supported by numerical simulations of photoemission in the quantum well potential system.
Figure \ref{Fig: Photoemission_1d} compares the photoelectron amplitudes calculated from the time-dependent first-order perturbation method, matrix element using the single-wave state, and that using the TR-LEED state.
It shows that the TR-LEED successfully reproduces the photoelectron amplitude directly calculated from the perturbation theory, while the single-wave state exhibits a slightly different behavior.
However, the difference in the simulated amplitudes between the TR-LEED and single-wave state is not large because the diffraction term is insignificant, showing that the single-wave state can be used to approximate the TR-LEED state.

Another type is obtained by modulating a plane wave by atomic potentials [17--19].
Final state wave functions are calculated around each atom and are multiplied by the phase shift to achieve wave-like coherence.
However, it is unclear how such a wave function shapes in a vacuum and whether it satisfies the boundary condition at the vacuum.

\section{Note 2: Magnetotransport properties}
\begin{figure}[ht]
\includegraphics{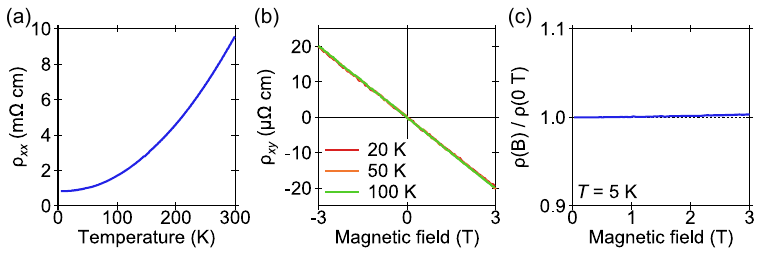}
\caption{\label{Fig: Transport} Magnetotransport properties of $1T$-$\mathrm{Ti}\mathrm{S}_2$. (a) Temperature dependence of the resistivity. (b) Magnetic field dependence of the Hall resistivity for three temperatures. (c) Magnetoresistance.}
\end{figure}

Magnetotransport measurements were performed using Quantum Design PPMS.
Figure \ref{Fig: Transport} summarizes the magnetotransport properties of $1T$-$\mathrm{Ti}\mathrm{S}_2$.
All the measured data exhibit trivial behavior, indicating the simple electronic structure of this material.

\section{Note 3: X-ray photoemission spectroscopy (XPS) measurements}
\begin{figure}
\includegraphics{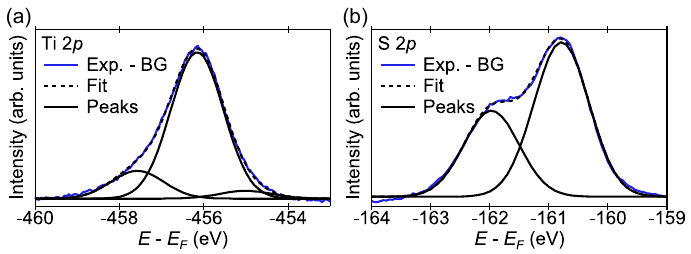}
\caption{\label{Fig: XPS} XPS spectra of (a) Ti $2p$ and (b) S $2p$. The blue curves represent the experimental spectra after the Shirley background [31] subtraction.}
\end{figure}

We performed XPS measurements using the Al $K\alpha$ line ($h\nu=1486.6\ \mathrm{eV}$).
Figure \ref{Fig: XPS} summarizes XPS measurement results of $1T$-$\mathrm{Ti}\mathrm{S}_2$.
We corrected the Fermi level so that the main Ti $2p_{3/2}$ and S $2p_{3/2}$ peaks appear at the same positions as the previous study [32].
The peak positions determined by the curve fitting were $-457.6$, $-456.2$, and $-455.0$ eV for Ti and $-162.0$ and $-160.8$ eV for S.

The previous study [32] reports the TiS structure can form due to sulfur escaping, and additional peaks related to the reconstructed structure appear at $-455.4$ eV for Ti and $-161.2$ eV for S.
Since they are absent in our XPS spectra, we conclude surface reconstruction does not happen in our samples.

\section{Note 4: Laue back-reflection measurements}
\begin{figure}[hb]
\includegraphics[width=100mm]{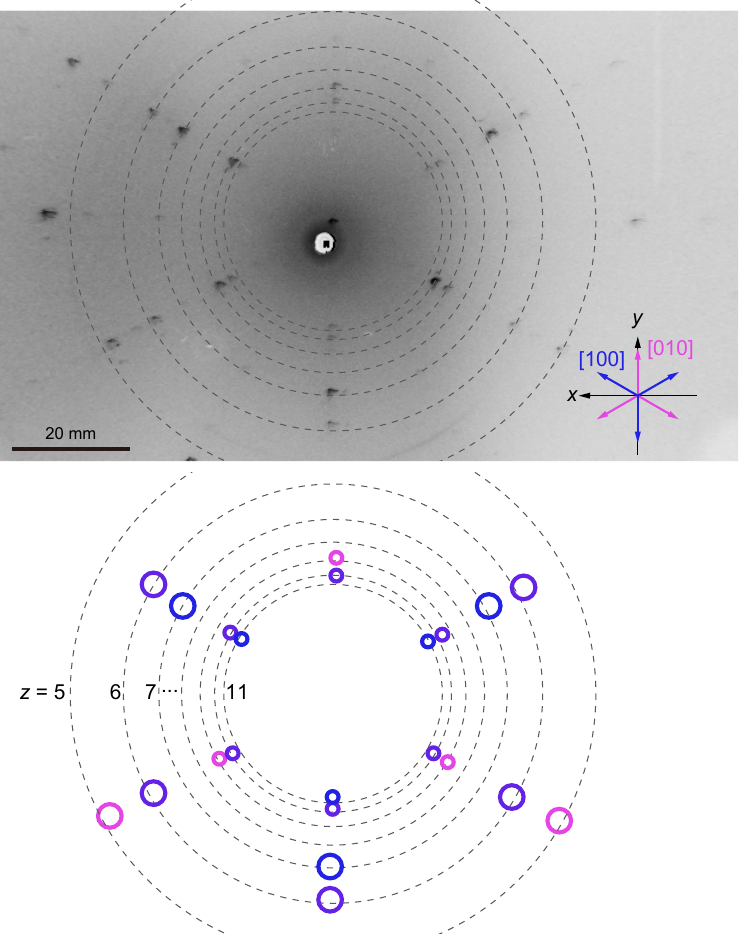}
\caption{\label{Fig: Laue} Laue image and indices for intense spots. The blue, pink, and purple circles represent peaks where $[10z]$ is stronger than $[01z]$, $[01z]$ is stronger than $[10z]$, and $[10z]$ is as strong as $[01z]$, respectively. }
\end{figure}

Laue back-reflection measurements were performed using a Laue camera system (IPX-LC, IPX Co. Ltd.) and a tungsten anode x-ray tube.
The tube potential was set to 30 keV, and the distance between the sample and the screen was 50 mm.
The three-fold rotational Laue pattern and absence of six-fold rotation in Fig.\ \ref{Fig: Laue} indicate the high quality of our crystals without rotational domains.
The $[100]$ and $[010]$ directions were distinguished by calculating the reflection intensity as represented by colored circles in Fig.\ \ref{Fig: Laue}; we used the atomic scattering factors tabulated in Ref.\ [33].

\clearpage

\section{Note 5: ARPES experiments}
ARPES measurements using a 7 eV laser and synchrotron radiation ($h\nu=9\text{--}39\ \mathrm{eV}$) were performed at ISSP, the University of Tokyo [53, 54] and HiSOR BL-9A, respectively.
This energy range, enabling us to measure the whole Brillouin zone with high energy resolution, has been frequently used in ARPES measurements.
The measurement temperature was around 50 K, and the energy resolution was 13 meV for laser ARPES and 30--40 meV for synchrotron ARPES.
Before the ARPES measurements, the samples were cleaved \textit{in situ} in ultrahigh vacuum better than $5\times10^{-7}\ \mathrm{Pa}$ to obtain clean surfaces.

\section{Note 6: Band dispersion calculations}
We used OpenMX [34]  and Quantum ESPRESSO [35--37] for band calculations using the Perdew-Burke-Ernzerhof (PBE) functional [38, 39] and Heyd-Scuseria-Ernzerhof (HSE) hybrid functional [40, 41], respectively.
OpenMX uses localized pseudoatomic orbitals as basis sets, and the linear combination coefficients are used in our photoemission intensity calculations.
However, since hybrid functionals are not implemented in OpenMX, we instead used Quantum ESPRESSO.

In PBE calculations, we used a 40-layer slab system, which is also used in the photoemission intensity calculations.
On the other hand, we employed a bulk $1T$-$\mathrm{Ti}\mathrm{S}_2$ in HSE self-consistent calculations to reduce computational costs.
After that, we used Wannier90 [42] to plot band dispersions projected on the $(001)$ plane.
The bulk unit cell size was fixed to the experimental one $a=3.405\ \text{\AA},\ c=5.695\ \text{\AA}$, and the electron spin was neglected in the calculations. 

\begin{figure}[htb]
\includegraphics{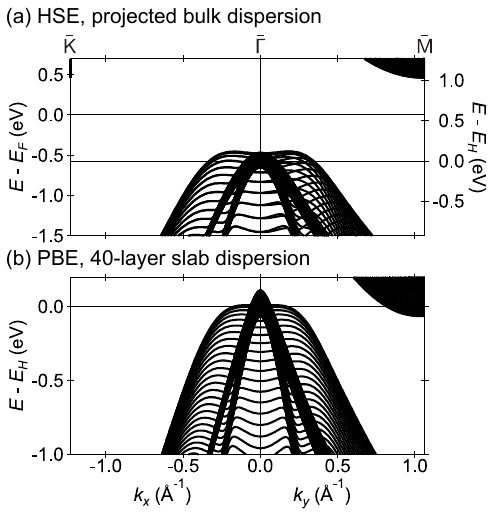}
\caption{\label{Fig: Band} Band dispersions of $1T$-$\mathrm{Ti}\mathrm{S}_2$. (a) Bulk band dispersion projected on the $(001)$ plane. The HSE hybrid functional was used. (b) Band dispersion of a 40-layer slab calculated. The PBE functional was used.}
\end{figure}

As represented in Fig.\ \ref{Fig: Band}, quasi-two-dimensional (weakly $k_z$-dispersing) hole bands and electron bands exist around the $\bar{\Gamma}$ and $\bar{M}$ points, respectively.
There is an indirect band gap between the hole and electron bands in the HSE band dispersion, while the gap is underestimated in the PBE band dispersion.
This behavior is consistent with the previous calculation study of this material [43].
The band gap underestimation frequently happens in PBE calculations, and hybrid functionals can achieve better coincidence with experimental band dispersions [44].

\section{Note 7: Surface relaxation}
We confirmed the surface relaxation effect is negligible for $1T$-$\mathrm{Ti}\mathrm{S}_2$ by the following procedure; we used OpenMX for these calculations.
First, we obtained the optimized bulk cell size for the PBE functional; $a=3.398\ \text{\AA},\ c=6.152\ \text{\AA}$, slightly larger than the experimental one, was obtained.
After that, we performed atom position optimization calculations using a 6-layer slab, where atoms in the bottom three layers were fixed, and those in the top three layers were free.
The displacement of atoms was at most 0.006 \AA, so we conclude that we don't need to consider the surface relaxation effect for this material.

\section{Note 8: FP wave function calculations}
\begin{figure}
\includegraphics{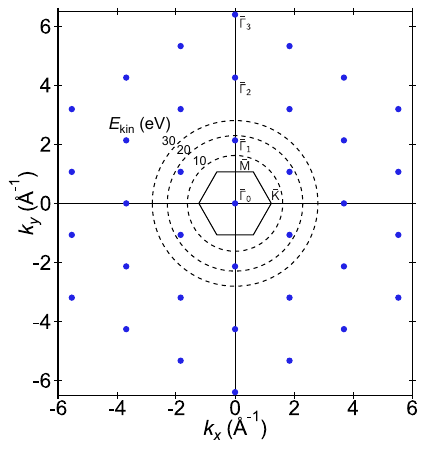}
\caption{\label{Fig: BZ_wide} Wide view of the reciprocal space. The blue dots represent reciprocal lattice points.}
\end{figure}

We calculated $\psi_{\mathbf{g}_\para}(z)$ for all the blue dots in Fig.\ \ref{Fig: BZ_wide} representing the in-plane reciprocal lattice points.
After wave function calculations, the decay function is multiplied to effectively include the surface-sensitive property of photoemission spectroscopy [55] because the Kohn-Sham system cannot derive the final state decay due to scattering between electrons.
While the nonzero $\kappa_{nz}$ component makes a decaying wave function, it does not properly reflect the surface sensitivity of ARPES.

The linear decay function $p(z)=\mathrm{max}(1-z/\lambda, 0)$ was employed, where $p(z)$ corresponds to the emission probability and $z$ represents the distance from the surface.
$\lambda$ can be interpreted as the probing depth of ARPES, while the calculation results presented below did not strongly depend on $\lambda$ [Fig.\ \ref{Fig: PAD_properties}].
In the main text, we show the $\lambda=4c$ results.

\section{Note 9: When the complex $z$ component is necessary}
\begin{figure}
\includegraphics{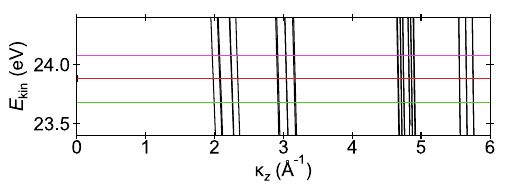}
\caption{\label{Fig: Band_kappaz} Bulk band dispersion of unoccupied states along the $\kappa_z$ axis. The in-plane momentum is $(0.1, 0.0)\ [\text{\AA}^{-1}]$, and the colored horizontal lines correspond to ones in Fig.\ 2(b). Only the real eigenvalues are plotted.}
\end{figure}

In the main text, we argued that the complex $z$  component is necessary when the kinetic energy is between the hybridization gap.
The other case appears on the imaginary axis ($k_z=0$).
First, when we approximate the crystal potential to be a constant $V_0$, a bulk eigenstate satisfies $\psi_{n\mathbf{g}}=1$ for a specific $\mathbf{g}$ and $\psi_{n\mathbf{g}^\prime}=0$ for the others.
The eigenenergy (in the Hartree atomic unit) is
\begin{equation}
\frac{1}{2}|\mathbf{k}+\mathbf{g}|^2+V_0=\frac{1}{2}\Bigl[|\mathbf{k}_\para+\mathbf{g}_\para|^2+(k_z+g_z-i\kappa_z)^2\Bigr]+V_0,
\end{equation}
so it becomes real when either $\kappa_z=0$ or $k_z=g_z=0$ is satisfied.
Since a complex $\mathbf{k}$ changes the Hamiltonian to be non-Hermite, not all eigenvalues are real.
The former is the real $z$ component case [Fig.\ 2(b)].
In the latter case, the eigenvalue is
\begin{equation}
\frac{1}{2}\Bigl[|\mathbf{k}_\para+\mathbf{g}_\para|^2-\kappa_z^2\Bigr]+V_0
\end{equation}
and the hole band is formed along the $\kappa_z$ axis.
The eigenvalue remains real even if the potential is adiabatically changed to the crystal one.
Therefore, it equals the specified kinetic energy when $\mathbf{g}_\para$ and $\kappa_z$ are relatively large [Fig.\ \ref{Fig: Band_kappaz}].
However, the linear combination coefficient $p_n$ for such a state is small because $\psi_{\mathbf{g}_\para}(z)$ for a large $\mathbf{g}_\para$ has to be connected to zero in a vacuum.

\begin{figure}
\includegraphics{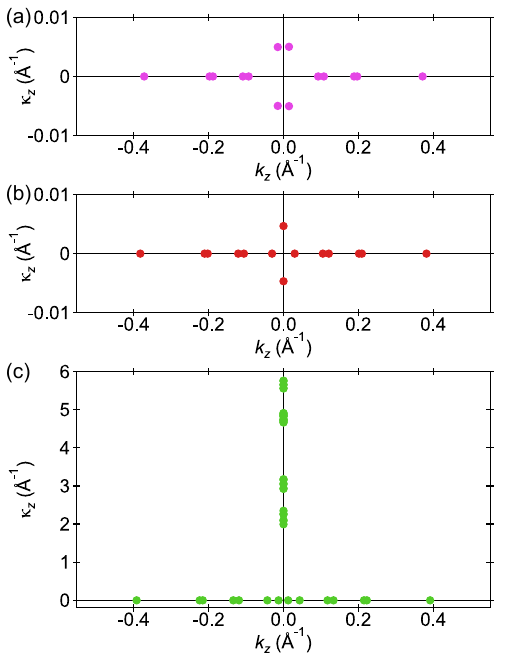}
\caption{\label{Fig: kz_kappaz_map} Distribution of the $z$ components with the specified kinetic energy. The colors of the dots correspond to the horizontal lines in Fig.\ 2(b). In (a) and (b), some dots in the large $\kappa_z$ region are not plotted.}
\end{figure}

Figure \ref{Fig: kz_kappaz_map} shows the search result.
The imaginary part due to the hybridization gap is relatively small [Figs.\ \ref{Fig: kz_kappaz_map}(a) and \ref{Fig: kz_kappaz_map}(b)], while that due to the hole band on the $\kappa_z$ axis is relatively large [Fig.\ \ref{Fig: kz_kappaz_map}(c)].

\clearpage

\section{Note 10: Calculation stability}
The Kohn-Sham system in OpenMX includes the local potential $V(\mathbf{r})$ and the projector-type non-local potential:
\begin{equation}
\hat{V}^\mathrm{nonloc}=\sum_p \lambda_p |\beta_p\rangle \langle \beta_p|.
\end{equation}
The projection operator $|\beta_{p}\rangle$ has nonzero values within the pseudopotential cutoff.
We did not include the non-local potential for the following reasons.
First, the Hamiltonian matrix element for the non-local potential depends on the bulk wavevector, so calculating it for every eigenvalue calculation is computationally heavy.
Second, the FP final state calculations with the non-local potential were unstable because we frequently failed to find an appropriate number of eigenstates.
The instability may be resolved by increasing the $\mathbf{g}_\para$ vectors to represent localized changes around the cores by the non-local potential, but it will make the computational cost unrealistically heavy.
We note that the FP wave function in the valence region outside the pseudopotential cutoff is accurate because it is determined by the boundary condition at a vacuum.

Another stability issue is the wave function of the top layer and a vacuum above the surface when the photon energy is smaller than 25 eV.
Since the circle determined by the photoelectron kinetic energy [Fig.\ 1(b)] does not enclose the $\bar{\Gamma}$ point, $\psi_{\bar{\Gamma}_1}(z)$ is of the form $e^{\pm\kappa z}$, not the form $e^{\pm ikz}$.
The real exponential functions easily diverge and cause the calculation instability.
The solution is to limit the $\mathbf{g}_\para$ within the circle and to exclude the exponential function from the wave function.
We applied this modification to calculate the photoemission intensity below 25 eV.

\clearpage

\section{Note 11: Synchrotron ARPES measurements}
\begin{figure}[hb]
\includegraphics[width=160mm]{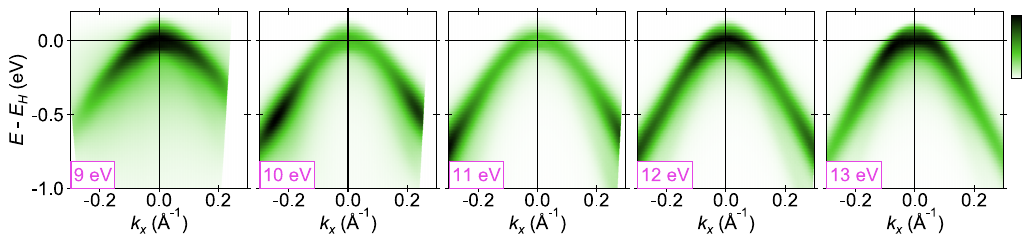}
\includegraphics[width=160mm]{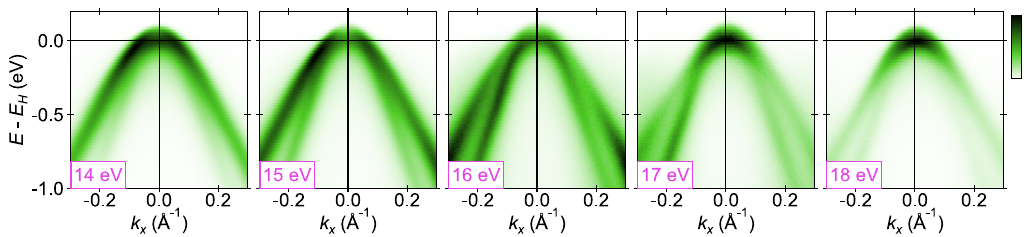}
\includegraphics[width=160mm]{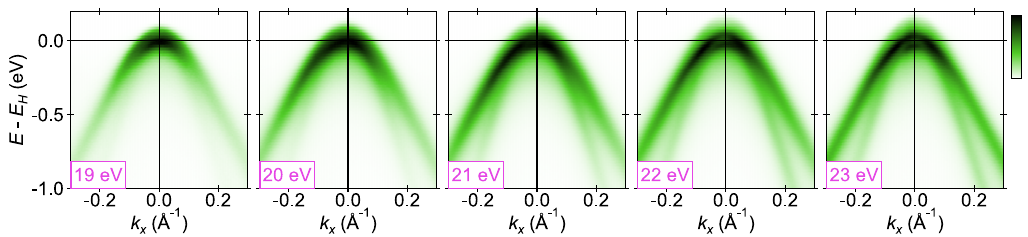}
\includegraphics[width=160mm]{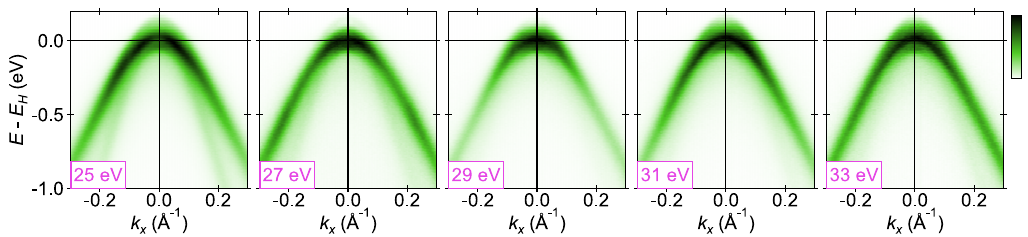}
\includegraphics[width=160mm]{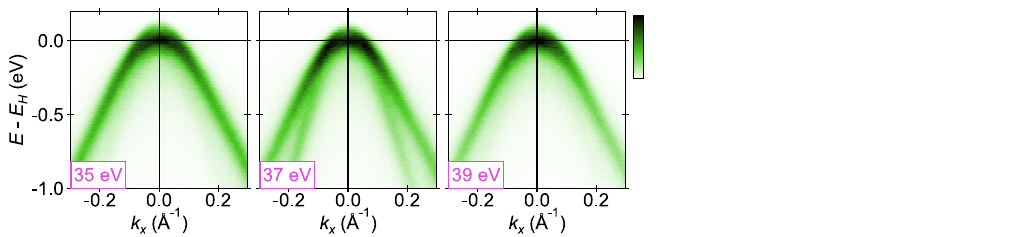}
\caption{\label{Fig: Eks_s} Photon energy dependence of synchrotron ARPES spectra using $s$-polarized light.}
\end{figure}

\begin{figure}[ht]
\includegraphics[width=160mm]{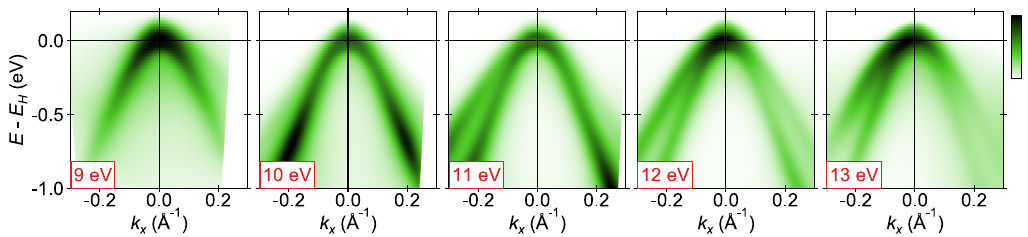}
\includegraphics[width=160mm]{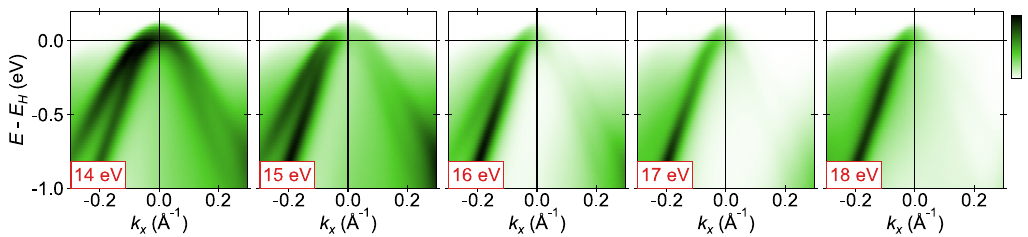}
\includegraphics[width=160mm]{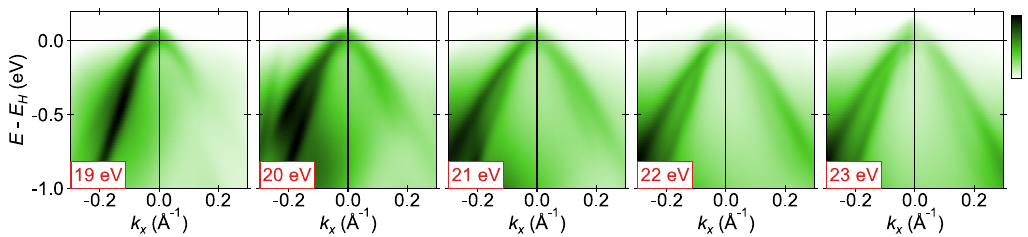}
\includegraphics[width=160mm]{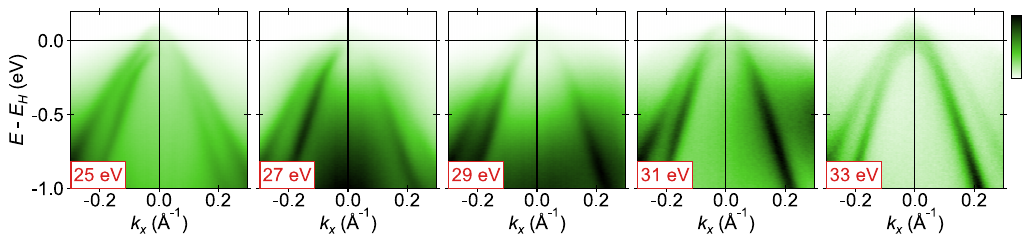}
\includegraphics[width=160mm]{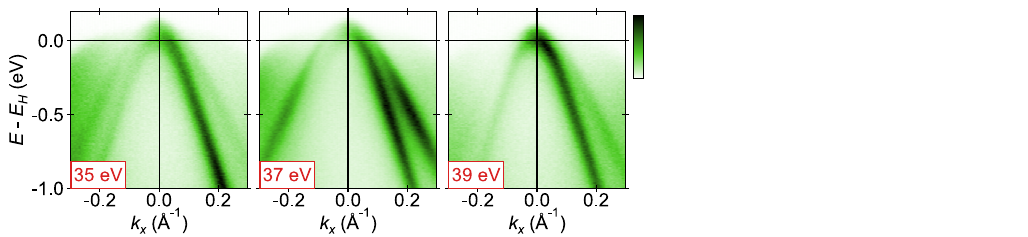}
\caption{\label{Fig: Eks_p} Photon energy dependence of synchrotron ARPES spectra using $p$-polarized light.}
\end{figure}

Figures \ref{Fig: Eks_s} and \ref{Fig: Eks_p} show the photon-energy and polarization dependence measurement data in our synchrotron ARPES measurements.

\clearpage

\section{Note 12: Momentum dependence of the asymmetry}
\begin{figure}[hb]
\includegraphics{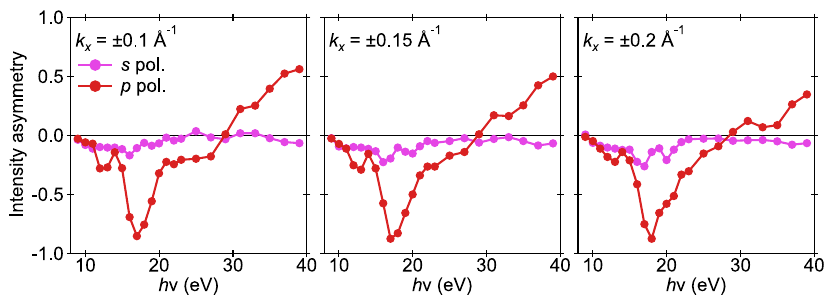}
\caption{\label{Fig: Asym} Photon energy dependence of the photoemission intensity asymmetry for $k_x=\pm0.1\ \text{\AA}^{-1},\pm 0.15\ \text{\AA}^{-1},\pm0.2\ \text{\AA}^{-1}$.}
\end{figure}

Figure \ref{Fig: Asym} shows the photon-energy-dependent intensity asymmetry calculated for three momenta.
The overall trend for the $p$-polarized spectra is the same between these three graphs.

\section{Note 13: Probing depth dependence of photoemission calculations}
\begin{figure}[hb]
\includegraphics{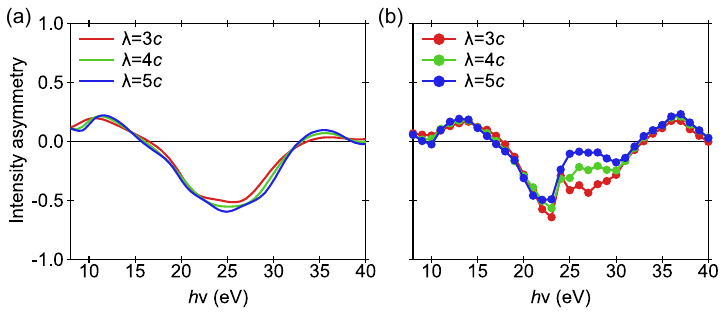}
\caption{\label{Fig: PAD_properties} Probing depth ($\lambda$) dependence of the calculated photoemission intensity asymmetry for (a) PW final states and (b) FP final states.}
\end{figure}

Figure \ref{Fig: PAD_properties} shows that the probing depth does not greatly affect the integrated intensity asymmetry of simulated spectra.

\section{Note 14: Laser ARPES measurements and photoemission calculations}
\begin{figure}[hb]
\includegraphics{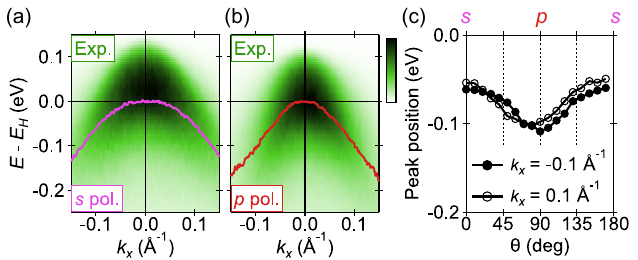}
\caption{\label{Fig: ARPES_laser} Laser ARPES measurement results. (a), (b) ARPES spectra for $s$- and $p$-polarized lights. (c) $\theta$ dependence of the peak positions at $k_x=\pm0.1\ \text{\AA}^{-1}$.}
\end{figure}

Figure \ref{Fig: ARPES_laser} summarizes our laser ARPES measurements.
While ARPES spectra were symmetric even in the $p$-polarization measurements, we observed peak position oscillation as a function of the polarization angle $\theta$.

\begin{figure}[ht]
\includegraphics{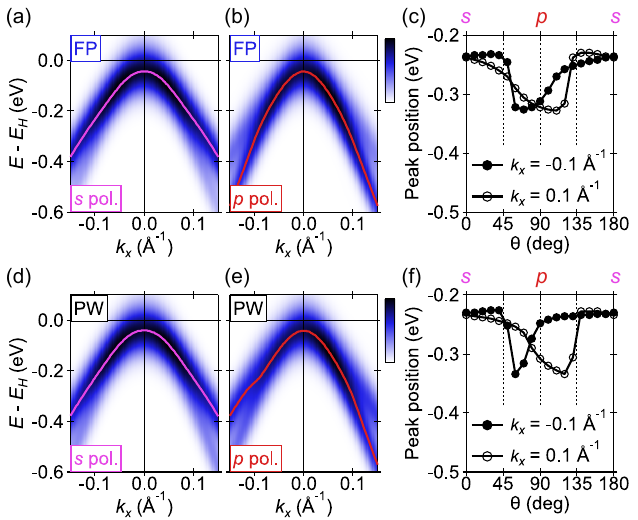}
\caption{\label{Fig: PAD_7eV} Photoemission intensity calculations for the 7 eV excitation. (a), (b) Simulated photoemission angular distributions for $s$- and $p$-polarized lights using the FP final states, respectively. The overlaid curves represent peak positions extracted from energy distribution curves. (c) $\theta$ dependence of the peak positions at $k_x=\pm0.1\ \text{\AA}^{-1}$. (d)--(f) Those for the PW final states.}
\end{figure}

Figure \ref{Fig: PAD_7eV} clearly shows that the FP final states reproduce experimental results [Fig.\ \ref{Fig: ARPES_laser}] better than the PW final states.
In our laser ARPES simulations with the $\mathbf{g}_\para$-limited condition, only the $\mathbf{g}_\para=0$ component was considered.
The $\mathbf{g}_\para\not=0$ components, excluded from our calculations, can substantially change the simulated ARPES spectra and contribute to better coincidence.

\clearpage
\section{Note 15: Amplitude analysis of the FP wave functions}
\begin{figure}[ht]
\includegraphics{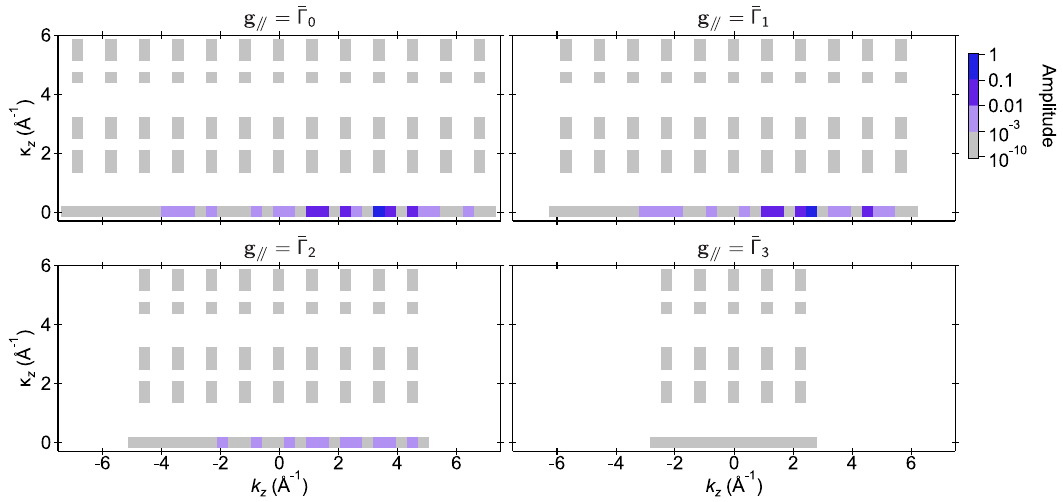}
\caption{\label{Fig: Components_map} Amplitude map of the FP wave function for $\mathbf{g}_\para=\bar{\Gamma}_0\text{--}\bar{\Gamma}_3$ [defined in Fig.\ \ref{Fig: BZ_wide}]. The other conditions are mentioned in the main text.}
\end{figure}

\begin{figure}[ht]
\includegraphics{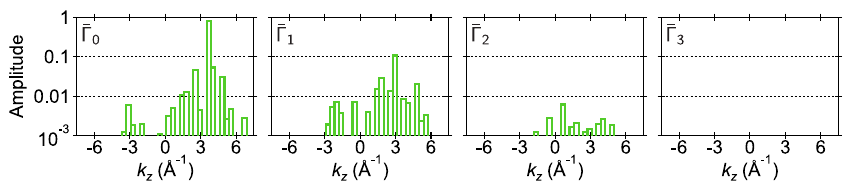}
\caption{\label{Fig: kz_components_map} Amplitude histogram of the FP wave function for $\mathbf{g}_\para=\bar{\Gamma}_0\text{--}\bar{\Gamma}_3$.}
\end{figure}

Figures \ref{Fig: Components_map} and \ref{Fig: kz_components_map} show that $\kappa_z\not=0$ waves and large $\mathbf{g}_\para$ components have only negligible amplitudes, although they are necessary to construct an appropriate basis set.

